\documentclass[showpacs,aps, prl, preprint]{revtex4}
\usepackage{graphics}
\usepackage{epsfig}
\begin{document}

\title{Astrophysical $S_{17}$(0) factor from a measurement of
$d(^7$Be,$^8$B)$n$ reaction at $E_{c.m.}$ = 4.5 MeV}

\author{J.J.~Das$^1$,    V.M.~Datar$^2$, P.~Sugathan$^1$,
N.~Madhavan$^1$, P.V.~Madhusudhana~Rao$^3$,   A.~Jhingan$^1$,
A.~Navin$^2$,    S.K.~Dhiman$^4$,   S.~Barua$^5$, S.~Nath$^1$,
T.~Varughese$^1$, A.K.~Sinha$^6$,   R.~Singh$^7$,    A.~Ray$^8$,
D.L.~Sastry$^3$, R.G.~Kulkarni$^9$, and R.~Shyam$^{10}$}
\affiliation{
$^1$Nuclear Science  Centre, P.O Box 10502, Aruna Asaf Ali Marg, New Delhi-
110067, India\\
$^2$Nuclear Physics Division, Bhabha Atomic Research Centre, Trombay, Mumbai-
400085, India\\
$^3$Department  of Nuclear  Physics,  Andhra  University,  Visakhapatnam-
530003,  India\\
$^4$Department of Physics, Himachal Pradesh University, Shimla-171005, India\\
$^5$Department of Physics, Gauhati University, Jalukbari-Guwahati-787014, 
India\\
$^6$IUC-DAEF Kolkata Centre, Bidhan Nagar,   Kolkata-700092,  India\\
$^7$Department of Physics and Astrophysics, Delhi University, New Delhi 
110007, India\\
$^8$Variable Energy  Cyclotron Centre, 1/AF  Bidhan Nagar, Kolkata-700064, 
India\\
$^9$Department of Physics, Saurashtra University, Rajkot-360005, India\\
$^{10}$Saha Institute of Nuclear Physics, 1/AF Bidhan Nagar, Kolkata-700064, 
India}

\date{\today}

\begin{abstract}

Angular   distribution  measurements  of  $^2$H($^7$Be,$^7$Be)$^2$H  and
$^2$H($^7$Be,$^8$B)$n$   reactions   at   $E_{c.m.}\sim$~4.5~MeV    were
performed  to  extract  the  astrophysical  $S_{17}(0)$ factor using the
asymptotic normalization coefficient (ANC) method. For  this  purpose  a
pure,  low  emittance  $^7$Be beam was separated from the primary $^7$Li
beam by a recoil mass spectrometer operated in  a  novel  mode.  A  beam
stopper  at  0$^{\circ}$  allowed  the  use  of  a  higher  $^7$Be  beam
intensity. Measurement of the elastic scattering in the entrance channel
using  kinematic  coincidence,  facilitated  the  determination  of  the
optical  model  parameters needed for the analysis of the transfer data.
The present measurement significantly reduces errors in  the  extracted
$^7$Be(p,$\gamma$)  cross  section  using the ANC method. We get   
$S_{17}$~(0)~=~20.7~$\pm$~2.4~eV~b.

\end{abstract}


\maketitle

Recently,  the  Sudbury Neutrino Observatory (SNO) group has discovered,
in a simultaneous measurement of the electron neutrino flux and the  sum
of  three  active  neutrino  fluxes,  that  a large fraction of the high
energy electron neutrinos emitted in the $\beta^+$ decay of $^8$B in the
Sun is transformed into other active neutrino flavors on  their  way  to
detectors  on the Earth~\cite{ahm02}. Together with the observation that
the reactor produced antineutrinos also oscillate~\cite{egu03}, it seems
that a solution of the solar neutrino problem is at hand.  To  ascertain
more accurately if there is a transformation of solar electron neutrinos
into  sterile  neutrinos the precision of both experimental measurements
and theoretical predictions must be improved~\cite{bah02}. In  order  to
make  more  accurate theoretical predictions of the $^8$B neutrino flux,
the  rate  [or  the  related  zero  energy   astrophysical   $S$-factor,
$S_{17}(0)$]  of  the reaction $^7$Be($p,\gamma)^8$B that produces $^8$B
in the Sun must be better determined~\cite{adl98}.

Recent  precision $^7$Be($p,\gamma$) measurements have yielded values of
$S_{17}(0)$  which  are  clustered  around  18.5~eV~b  \cite{ham01}  and
22.0~eV~b~\cite{jun03} and are not consistent with each other within the
quoted  errors.  In  view  of  this  discrepancy  the  determination  of
$S_{17}(0)$ by other  methods,  with  different  systematic  errors,  is
necessary.   One   kind   of   indirect  measurement  uses  the  Coulomb
dissociation of  $^8$B~\cite{mot03}  where  the  latest  experiment  has
yielded  a  $S_{17}(0)$  of  18.6~eV~b~\cite{sch03}. The other technique
uses the ($^7$Be,$^8$B) transfer reaction to extract  the  magnitude  of
the  asymptotic  radial  wave  function, characterized by the asymptotic
normalization coefficient (ANC), of the proton in $^8$B.  This  is  then
used  to  calculate the value of $S_{17}(0)$~\cite{xu94}. The ANC method
has been experimentally validated through the agreement of the  measured
$^{16}$O(p,$\gamma$)  cross  section  and  that derived from the data on
$^{16}$O($^3$He,d)$^{17}$F    reaction~\cite{gag99}.    Recently,    the
$S_{17}(0)$   has   been   extracted   from   the   transfer   reactions
$^{10}$B($^7$Be,$^8$B)$^9$Be and $^{14}$N($^7$Be,$^8$B)$^{13}$C yielding
values      of       17.8~$\pm$~2.8       and       16.6~$\pm$~1.9~eV~b,
respectively~\cite{azh99}.

Because  of  the  precise  knowledge  of the proton wave function in the
deuteron and a simpler exit channel (which makes  it  more  amenable  to
continuum discretized coupled channel~ (CDCC) methods \cite{oga03}), the
$d(^7$Be,$^8$B)$n$  reaction  may  still be an attractive choice for the
extraction of $S_{17}(0)$ with the ANC  method  provided  the  data  are
taken  at  beam  energies where this reaction is peripheral. It has been
shown~\cite{fer99} that for  entrance  channel  center  of  mass  (c.m.)
energies  around  or  below 6.0 MeV this condition is fulfilled for this
reaction. Liu {\it et al.}~\cite{liu96} have reported a  $S_{17}(0)$  of
27.4~$\pm$~4.4~eV  b  from  the  analysis  of  their  $d(^7$Be,$^8$B)$n$
experiment done at $E_{c.m.}=$~5.8~MeV. However,  it  has  been  pointed
out~\cite{gag98,fer99}  that  there  may  be  a large uncertainty in the
value of $S_{17}(0)$ extracted from this experiment due to lack  of  the
knowledge  of  the  entrance  and  exit channel optical model parameters
(OMP) as the corresponding elastic scattering data are not available.

The present work is a significant improvement over the earlier work of
Liu  {\it  et al.}~\cite{liu96}.   The  $^7$Be  beam  was produced  by
operating the existing recoil mass spectrometer, HIRA~\cite{sin94}, in
a novel  optical mode  leading to a  beam of superior  quality (purity
$>99.9  \%$, beam  spot  size $\approx  3$~mm,  angular divergence  of
$\pm$1$^{\circ}$). This allowed the use  of a stopper in the beam path
enabling the use  of a higher intensity (by  a factor of $\approx$~10)
beam  thus  reducing  the  statistical  errors  significantly.   In  a
separate experiment,  to obtain the entrance channel  OMP, the elastic
scattering   angular  distribution   was   measured  using   kinematic
coincidence to  eliminate the  background from other  target elements.
These have allowed a  more precise measurement of the d($^7$Be,$^8$B)n
angular distribution at the  lowest energy of $E_{c.m.}~=~4.5$~MeV and
the determination of $S_{17}(0)$ by the ANC method.

The measurements were made at Nuclear Science Centre, New Delhi, using
a  radioactive ion beam  of 21.0  $\pm$ 0.5~MeV  $^7$Be incident  on a
1.05~mg/cm$^2$  deuterated   polyethylene  (CD$_2$)$_n$  target.   The
$^7$Be  was   produced  through  the   p($^7$Li,$^7$Be)n  reaction  at
E($^7$Li)~=~25~MeV   using    a   pulsed   $^7$Li    beam   (intensity
$\sim$~3$\times$10$^{10}$/sec)  from the  15UD Pelletron,  at a  4 MHz
repetition  rate and FWHM  $\sim$2~nsec, to  obtain an  average $^7$Be
intensity of 3000/sec.  The  deterioration of the production target, a
20~$\mu$m polyethylene  (CH$_2$)$_n$ foil, was  minimized by automated
linear and  rotary motions.  The  forward going $^7$Be  particles were
selected using the  recoil mass spectrometer, HIRA, operated  in a new
ion    optical   mode   optimized    for   such    inverse   kinematic
reactions~\cite{das98}. In this mode the reaction products of interest
were  focused through  a slit  placed at  the center  of  the magnetic
dipole to reject the primary beam. The selected secondary ion beam was
refocused at  the target such that  the beam spot  was a
replica  of that  at  the production  target.  Two silicon  telescopes
placed at $\pm$  30$^{\circ}$ with respect to the  beam direction were
used in the primary target  chamber to measure the recoil protons. The
ratio of counts  in the recoil proton peak  from the production target
to  the  counts  of  $^7$Be  in the  secondary  reaction  chamber  was
monitored and  kept constant to  ensure an identical  and reproducible
trajectory of $^7$Be through HIRA. A 3~mm diameter graphite collimator
placed 86~mm upstream  of the target was used  to limit any unforeseen
wandering of  the beam spot.  The X-Y profile  of the $^7$Be  beam was
monitored  using  a  multi-wire  proportional  counter  (MWPC)  placed
$\sim$~90~mm behind the target. The MWPC had entrance and exit windows
of 1.5~$\mu$m polyethylene  and was operated using isobutane  gas at a
pressure of  5~mbar. A 4~mm diameter  tantalum disk mounted  on a thin
(0.25~mm  $\phi$) wire,  at  a  distance of  114~mm  from the  target,
stopped  the main beam.  This reduced  the beam  flux incident  on the
downstream detector  telescope by a  factor of $\sim$8.   The detector
telescope  consisted  of  a  $\Delta$E  gas  ionization  chamber  (IC)
followed by  a 50$\times$50~mm$^2$ two  dimensional position sensitive
Si-E detector~(PSSD).   The IC had an  opening of 55$\times$55~mm$^2$,
active depth of 60~mm, and was  operated at a pressure of 50~mbar. The
position and energy resolutions of the PSSD and IC were measured to be
$<$2~mm, 200~keV  and 200~keV, respectively, using  a $^{241}$Am alpha
source.  The  energy, position,  pileup parameter (generated  from the
zero crossover time  of the bipolar E pulse) and  time of flight (TOF)
with  respect to  the  timing  reference of  the  beam pulsing  system
together  with the scaled  down monitor  detector energy  signals were
recorded  in  an event  by  event mode.  The  MWPC  energy output  was
recorded independently in a multichannel analyzer to minimize the dead
time  in the  data acquisition  system. This  was used  to  obtain the
integrated $^7$Be beam intensity. A 10 Hz precision pulser was used to
monitor the  dead time of the  data acquisition system as  well as the
electronic gain of, and noise in, the detector system. The response of
the  detector  telescope  was  continuously  monitored  using  a  weak
$^{229}$Th  alpha source.   The count  rate in  the detector  was kept
constant  to within  $\pm$~10\%  so  as to  keep  the pileup  fraction
similar for different runs.

The beam  profile was maintained within $\pm$~0.25~mm  during the runs
for an accurate  angular definition.  The  stopper  allowed  a ten  fold
increase  in the  $^7$Be intensity  as compared  to that  of  Liu $\it
et~al$.~\cite{liu96} for  a similar pileup rate.  HIRA  was rotated to
2$^{\circ}$  and  the  scattered  $^7$Li,  $^7$Be and $^{12}$C ions were
selected to calibrate the particle identification (PID) of the  detector
telescope  in  situ.  This was essential in choosing the $\Delta$E-E two
dimensional~(2D) gates for $^8$B in conjunction with a SRIM~\cite{zei85}
calculation. The  (CD$_2$)$_n$  and  (CH$_2$)$_n$  data  with  (without)
stopper   were   taken   for   4$\times$10$^8$   (8$\times$10$^7$)   and
1.5$\times$10$^8$   (2$\times$~10$^7$)   $^7$Be   incident    particles,
respectively. The PID parameter was calculated in the standard way using
the  $\Delta$E-E  information~\cite{bromley}.  Typical PID spectra after
gating on the pileup, time-of-flight and particle energy (to remove  the
elastically  scattered  $^7$Be  from the target) are shown in Fig.~1. The
$^8$B events were selected using suitable gates on PID, pileup parameter
and TOF. The  angular  distribution  was  obtained  using  the  position
information  from  the  PSSD.  The  position  response of the latter was
measured offline using  an  alpha  source  and  a  mask  placed  on  the
detector.  An  overall angular resolution of 0.9$^{\circ}$ was estimated
taking into account the angular  divergence,  transverse  beam  profile,
angular straggling in the target and gas, and the position resolution of
the PSSD. The $^8$B yields were corrected for small dead time losses and
transmission loss in the MWPC.

Elastic  scattering  for  the  $^7$Be  + $d$ system was also measured at
20.3$\pm$0.5~MeV using the same (CD$_2$)$_n$ target.  The  schematic  of
the  experimental  setup  is  shown in Fig.~2a. In an elastic scattering
event, the recoiling deuteron is detected in the annular  detector  (A2)
and  $^7$Be  in the forward detector telescope $\Delta$E(gas)-Si 2D. The
$^7$Be  particles  reaching  the  detector  telescope  as  a  result  of
scattering from the collimator and $^{12}$C~(in (CD$_2$)$_n$ target) and
from  the beam halo have no coincident events in A2 and were eliminated.
Although the inelastic contribution from the 429  keV  state  in  $^7$Be
could not be resolved, its contamination to the elastic scattering yield
is  expected  to  be small (also supported by low energy $^7$Li+$^{12}$C
scattering  measurements)~\cite{kai03}.  This  has   been   checked   by
performing  a  calculation for the inelastic excitation cross section to
the first excited state in $^7$Be as discussed below. Fig.~2b shows  the
experimental elastic differential cross-sections.

These  data were  fitted in  a standard
optical  model  analysis using  the  code SNOOPY8Q~\cite{sno80}.   The
depths of the real and imaginary parts of the $d$-OMP were constrained
by  the method described  in Ref.~\cite{wal76}  where the  $d$-OMP was
calculated by folding the  nucleon optical potentials corresponding to
half the deuteron  energy so as to resolve  the discrete ambiguity in
the  $d$-OMP.  Following Ref.~\cite{bin71},  the radius  parameters of
both real and  imaginary parts were varied in the range  of 3.5 to 4.5
fm. In the fitting procedure, we applied an additional constraint that
the total reaction cross  section ($\sigma_R$) be close to $\sim$~1.2 b
which   is  obtained   by  calculations  done   with  the   OMP  of
Ref.~\cite{bin71} for  the $d$+$^7$Li system at  comparable energies and
also  from the measurements  of $\sigma_R$  reported for  the $d+^9$Be
system~\cite{auc96}. Four sets of  best fit potentials obtained from a
$\chi^2$ minimization analysis of the  data are shown in Table I. The fit
to the  elastic angular distributions obtained with  potential S1 is shown
in Fig.~2b (dashed line). Also shown in this figure is the sum of the
elastic and inelastic (to the first excited state in $^7$Be at 427 keV)
cross sections (solid line). The latter has been calculated using the same
set of d-$^7$Be OMP and $\beta_2$ of 0.6~\cite{nunes97}. Results obtained
with sets S2~-~S4 are similar and cannot be distinguished  from these 
curves. This figure also shows that the potentials sets 1 and 2 of 
Ref.~\cite{liu96} provide very poor fits to our elastic data. Similar
poor fits are obtained using the potential sets given in
Refs.~\cite{gag98,fer99}. While the fits could be improved upon by
performing measurements with better statistics and having more data
points, the present data on elastic scattering angular distribution
can clearly discriminate between the different deuteron optical potentials
and the potential sets extracted by us are the only ones which provide
any reasonable fit to these data. 

The  measured $d(^7$Be,$^8$B)$n$ angular distribution (shown in Fig.~3)
has been analyzed within the finite range distorted wave Born  
approximation (FRDWBA) using the code
DWUCK5~\cite{kunz}(with  full  transition  operator) which  has  been
modified to include  external  form  factors (FF).   The  FF for  the
deuteron-neutron  overlap has  been  obtained from  the deuteron  wave
function (including both $s$ and $d$ states) corresponding to the Reid
soft core potential.  A two body model for the $p$ - $^7$Be system has
been assumed where the proton  occupies a single particle state $n\ell
j$ and the $^8$B ($p$ -  $^7$Be) overlap function is written as ${\cal
S}^{1/2}  u_{n\ell j}(r)$.   Here $u_{n\ell  j}(r)$ is  the normalized
single  particle   radial  wave  function   and  ${\cal  S}$   is  the
spectroscopic factor which is directly related to the ANC~\cite{xu94}
and subsequently to the $S_{17}$(0) factor.
While the results  obtained with the proton in  the $0p_{3/2}$ orbital
are  given   here,  those  calculated  using   the  $0p_{1/2}$  proton
configuration separately are almost identical.  Five sets \cite{per76}
of the  neutron OMPs  were used in  the FRDWBA  transfer calculations.
These  were  obtained  from  the  global  parameterizations  given  in
Ref.~\cite{wil64} (used  extensively in Hauser-Feshbach calculations),
from fits to  $n$ + $^{10,11}$B scattering at 9.72  MeV (two sets) and
to $p  + ^9Be$  scattering at 5  and 6  MeV (two sets).   The compound
nuclear  (CN) contributions  were calculated  using  a Hauser-Feshbach
code  HAFEST   \cite{esw89}.   As  can  be  seen   from  Fig.~3, these
contributions   are   small   ($\approx$~7.5\%   at   44$^\circ$   and
$\approx$~1\% at 8$^\circ$).  The  uncertainty in the CN contributions
was estimated to be around 50$\%$ (which, however, adds only about 1\%
to the  overall theoretical uncertainty).   This has been  included in
the  systematic error  of  the extracted  $S_{17}$(0). The  calculated
transfer angular  distributions have been  folded using a  Monte Carlo
simulation  which took into  account the  spatial, angular  and energy
spread  of  the beam,  the  finite thickness  of  the  target and  the
position resolution  of the  detector.  The measured  transfer angular
distribution  below 45$^\circ$  was used  to extract  the $S_{17}$(0).
This was done to (a) minimize the error arising from the contributions
of  CN and  higher  order  processes which  may  affect the  extracted
$S_{17}(0)$ and (b)  use the forward angle data  to the maximum extent
in  order  to  reduce   the  statistical  errors.   These  theoretical
calculations at forward angles ($\theta_{c.m.}$ $\le$ 45$^\circ$) were
scaled  to  find  the  best  fit  to  the  data  (from  which  the  CN
contribution is subtracted) yielding ${\cal S}$ which has been used to
calculate    $S_{17}$(0)    by     the    procedure    discussed    in
Ref~\cite{xu94}. While calculating the $S$-factor the correction arising
from the $p-^7$Be scattering length has been included~\cite{baye2000}.

We have  verified that the peripheral condition of the transfer reaction
is fulfilled at our beam energy by two ways: (1) by using four different
sets of bound state potentials  for  the  $p$-$^7$Be  system  (given  in
Refs.~\cite{bar80,tom65,esb96,rob73})  which  lead to a variation of the
ANC by only $\pm$3.5$\%$ whereas the calculated transfer cross  sections
changed  by  about  30$\%$, and (2) by introducing a lower cutoff in the
radial integrals where it was found that results (at the forward angles)
obtained with a lower cutoff of up to 4.0 fm were  almost  identical  to
those  obtained  with  no lower cutoff. It should also be noted that the
multi step processes (inelastic excitation + transfer,  breakup  fusion)
and  core  excitation  in  $^8$B  have  negligible effect on these cross
sections~\cite{mor03}.

The  value  of  $S_{17}$(0)  and  the systematic errors arising from the
uncertainties in the $d$+$^7$Be OMP, $n$+$^8$B OMP  parameters  and  the
bound  state  wave  function  of the proton in $^8$B have been estimated
from 4, 5 and 4 choices, respectively, for each of these inputs. Each of
these  80  combinations  was  used  to  derive  the  best  fit  to   the
experimental  transfer  angular distribution and hence the corresponding
$S_{17}$(0). The range in which our calculated angular distributions lie
can be seen from Fig.~3 (short and long dashed  lines).  The  calculated
angular  distributions  agree well with the data within the experimental
uncertainty. The inset of Fig.~3 shows a histogram  for  the  number  of
occurrences/0.5~eV~b (N) for the $S_{17}$(0) which ranges from 18.8~eV~b
to  22.1~eV~b  with a calculated mean and RMS deviation of 20.7~eV~b and
0.9~eV~b, respectively. The distribution of  these  derived  $S$-factors
should  give  a  reasonable  estimate  of  the theoretical uncertainties
considering the large number of combinations for  the  potentials  used.
The  total  systematic  error  after including the uncertainty in target
thickness ($\pm$2\% by weighing samples from the same  stock  of  target
material)  is $\pm$~1.0~eV~b. The statistical error estimated from these
fits is $\pm$~1.4~eV~b. 

If the above exercise is carried out using the first 3, 5 and all 
11 data points, starting from the most forward angles, the extracted 
$S_{17}$(0) turns out to be 22.3$\pm$~2.7~eV~b, 22.7$\pm$~1.9~eV~b 
and 18.5$\pm$~1.6~eV~b, respectively. The 
mean S-factor from the analysis using these data sets and the 8 data point 
set used earlier is 20.6~eV~b. We may use this to estimate an additional   
error arising from the different choices of data points. This is 
probably a very conservative estimate of the error since it is expected 
that using the larger angle data makes it prone to contributions 
from higher order processes and uncertainties in compound 
nuclear contributions while use of only the 
most forward data points increases the statistical error while not 
making optimal use of the data. Nevertheless if this spread of 1.7~eV~b 
is added in quadrature to the statistical and systematic errors mentioned 
earlier the overall error increases to 2.4~eV~b. 
Since the systematic error alone contributes about 70$\%$ to the 
total error there is scope for reducing it. This would require a higher 
statistics elastic scattering and transfer measurement covering a larger 
angular range and the more elaborate CDCC calculations. 

In   conclusion,  we  have  measured,  for  the  first  time,  both  the
$d(^7$Be,$^8$B)$n$   transfer   and   the   entrance   channel   elastic
differential  cross  sections at the lowest beam energy hitherto using a
high quality $^7$Be beam from a recoil mass spectrometer operated  in  a
novel optical mode. The extracted $^7$Be(p,$\gamma$) $S_{17} (0)$-factor
is  20.7~$\pm$~2.4~eV~b is in good agreement with
the latest direct capture measurements~\cite{jun03} and those determined
from the CDCC analysis of  the  $^8$B  breakup  reaction~\cite{ogata05}.
Thus  the disagreement in the values of S$_{17}$(0) determined by direct
and indirect methods is reduced  (see  also  \cite{esb05}).
This experiment, therefore, clearly demonstrates that the ANC method can
be  used  for reasonably precise measurements of other (p,$\gamma$) 
$S$-factors involving short lived nuclei, where direct  capture measurements 
may be very difficult, if not impossible.

We  thank  G.K.  Mehta  for  his  valuable support and encouragement, D.
Beaumel  for  providing  the  (CD$_2$)$_n$  target,  Suresh  Kumar   for
providing the Hauser-Feshbach code, E.T. Subramaniam for support in data
acquisition and analysis and the NSC accelerator crew for delivering the
required  primary  beam.  We  also thank C.V.K. Baba, R.K. Bhowmik, D.R.
Chakrabarty, S.K. Datta, S. Kailas and A. Roy  for  useful  discussions.
Financial  support  from the Department of Science and Technology, Govt.
of India (grant no. SP/S2/K-26/1997) for development of the $^7$Be beam,
and from Department of Atomic Energy, Govt. of India to one of us  (SKD)
(sanction no. 2001/37/14/BRNS/699) are gratefully acknowledged.

\newpage
\begin{table}[here]
\caption{Parameters  of  the   Woods-Saxon  optical  model  potentials
extracted from the analysis of  the present $d$ + $^7$Be ($E_{c.m.}$ =
4.4 MeV)  elastic scattering data. A  spin orbit term  with $V_{so}$ =
8.60 MeV, $r_{so}$  = 2.17 fm, and $a_{so}$ = 0.61  fm, has been added
to all the potential sets. The optical potential is defined as that in
Ref. \protect~\cite{per76} with the light convention for the radius.}
\begin{ruledtabular}
\begin{tabular}{cccccccc}
Pot. & $V_0$ & $r_0$ & $a_0$ & $4W_s$ & $r_s$ & $a_s$ \\
    & \footnotesize (MeV) & \footnotesize (fm) & \footnotesize (fm) &
\footnotesize (MeV) & \footnotesize (fm) & \footnotesize (fm) \\
\hline
S1 & 103.12 & 2.23 & 0.62 & 79.03 & 2.37 & 0.17 \\
S2 & 107.87 & 2.17 & 0.61 & 58.84 & 2.28 & 0.25 \\
S3 &  92.54 & 2.41 & 0.57 & 117.50& 2.45 & 0.14 \\
S4 & 121.49 & 1.97 & 0.66 & 54.88 & 2.38 & 0.28 \\
\hline

\end{tabular}
\end{ruledtabular}
\end{table}

\newpage
\begin{figure}
\includegraphics[scale=1.0]{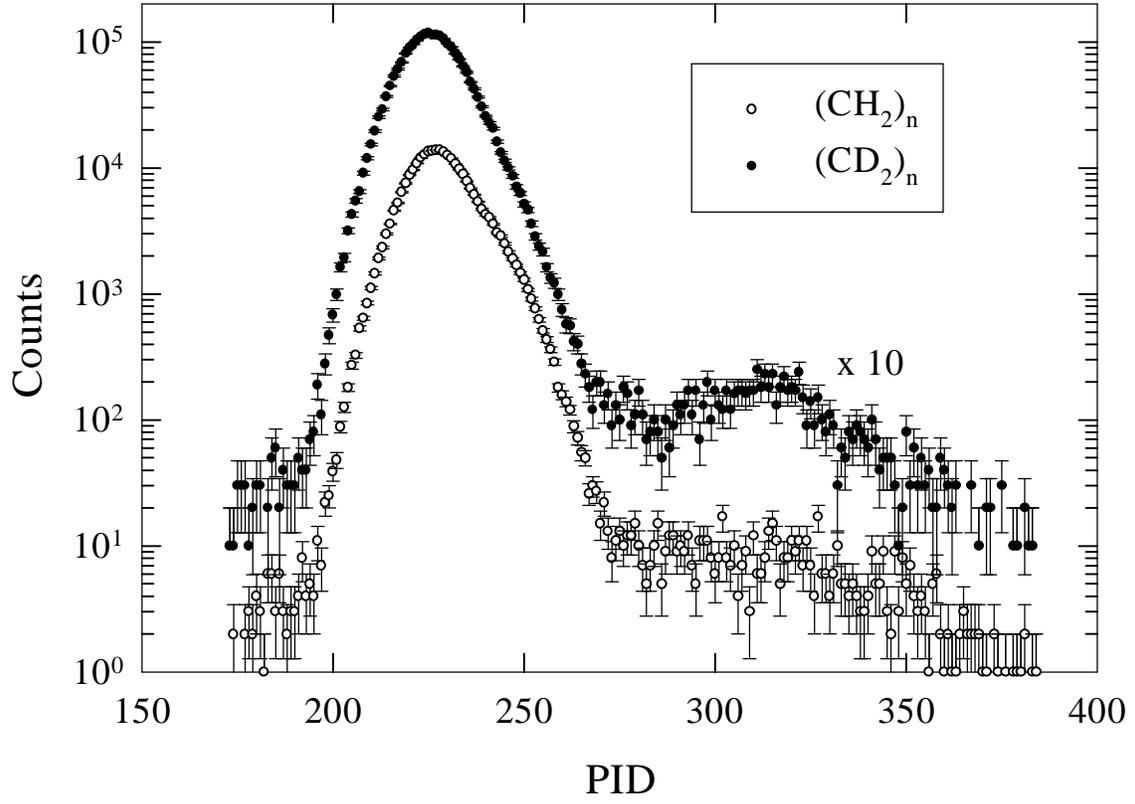}
\caption{a) Particle identification (PID) spectrum for the
polyethylene target with suitable energy, pileup and TOF cuts (see text)
for a integrated $^7$Be flux of 1.65$\times$10$^8$ particles.
(b) PID spectrum for deuterated polyethylene target.}
\end{figure}

\newpage
\begin{figure}
\includegraphics[scale=1.0]{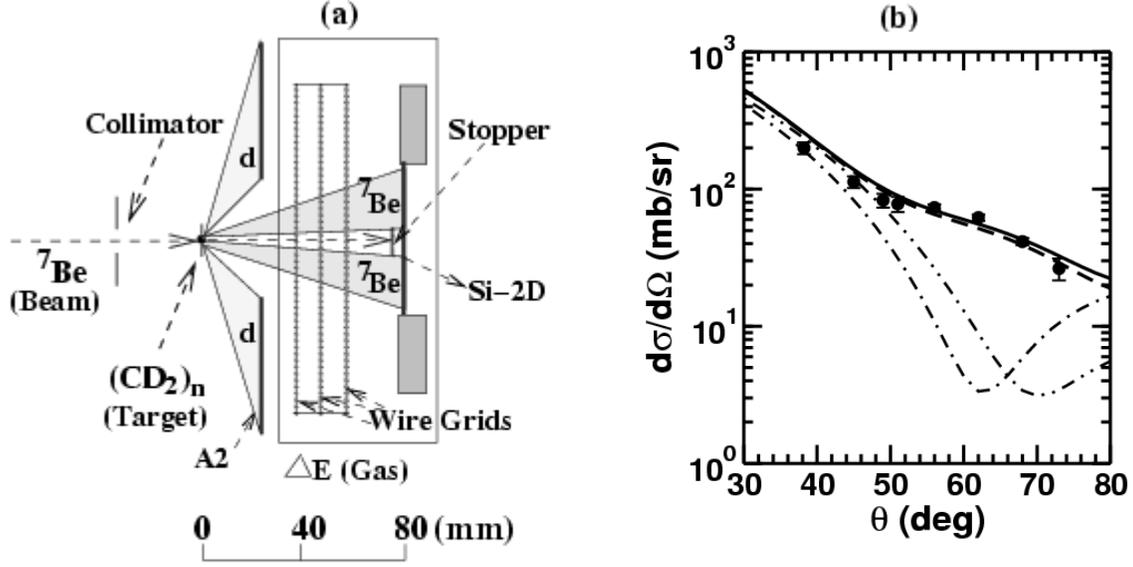}
\caption{a) Schematic  of detector  setup used for  elastic scattering
measurement.  b) Angular  distribution for  the elastic  scattering of
deuteron on  $^7$Be at E$_{c.m.}$ =  4.4 MeV.  Curves are  fits to the
data  using optical  potentials S1  (solid), and  S4(dotted)  of Table
I.  Results  obtained  from  potential  sets  of S2  and  S3  are  not
distinguishable  from those  shown.  The  dashed and  dot-dashed lines
represent the  cross sections  obtained with potential  sets 1  and 2,
respectively of Ref. ~\protect\cite{liu96}.}
\end{figure}

\newpage
\begin{figure}
\includegraphics[scale=1.0]{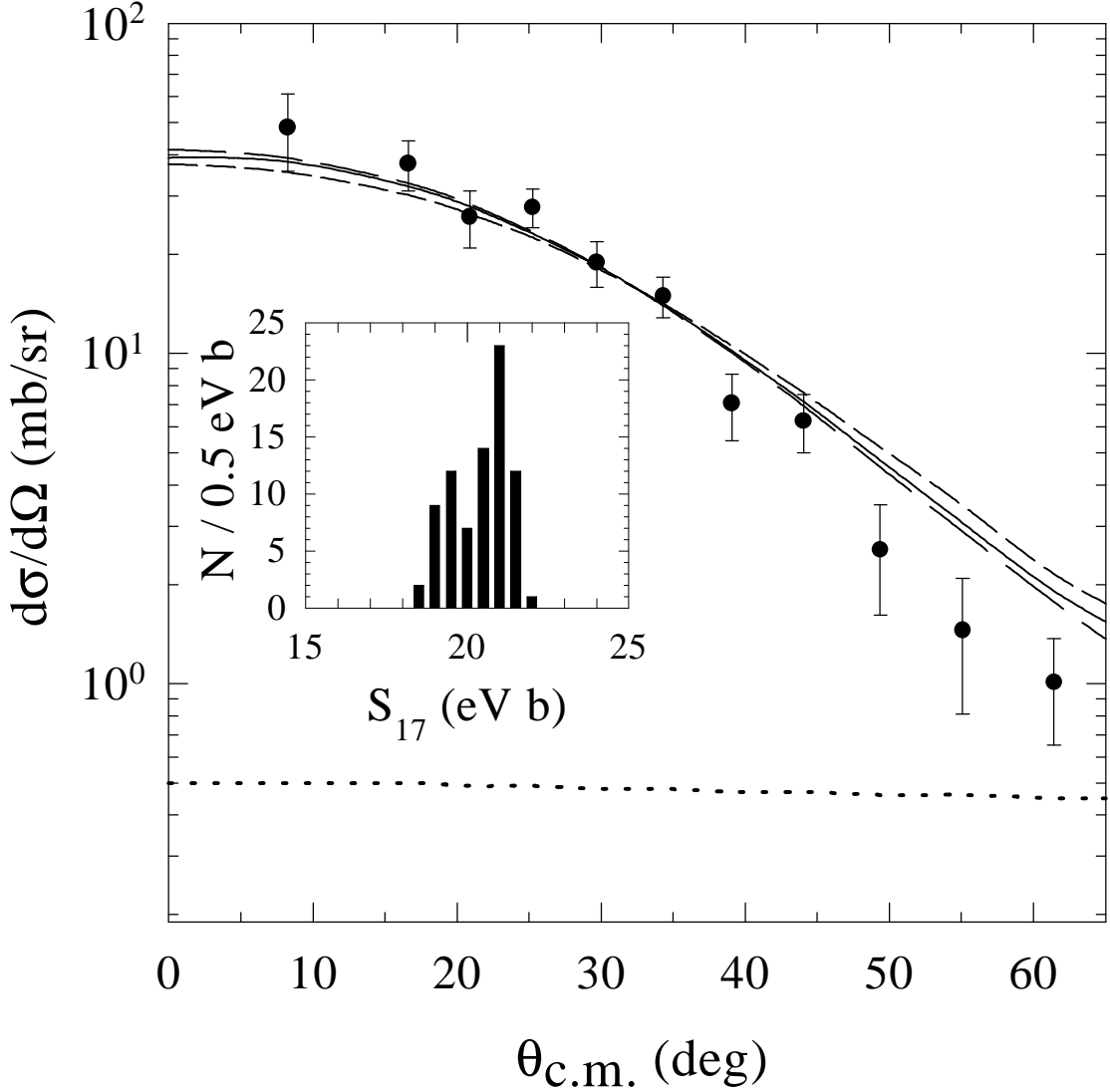}
\caption{Measured  $d(^7Be,^8B)n$ angular  distribution  together with
the  folded FRDWBA~$+$~CN  cross  sections shown  as  solid and short
and long dashed lines lines  (see text). The calculated compound nuclear
contributions are shown by the
dotted  line.  The  inset  shows  a histogram  plot  of the  extracted
S$_{17}$(0)  using the various  combinations of  $d$-$^7$Be, $n$-$^8$B
OMP and $p$-$^7$Be bound state potentials (see text).}
\end{figure}

\end{document}